\documentclass{epl}

\newcommand{\be}{\begin{equation}}      
\newcommand{\ee}{\end{equation}}      
\newcommand{\bea}{\begin{eqnarray}}      
\newcommand{\eea}{\end{eqnarray}}      
      
\newcommand{\bef}{\begin{figure}}      
\newcommand{\eef}{\end{figure}}

\def\spose#1{\hbox to 0pt{#1\hss}}       
\def\ltapprox{\mathrel{\spose{\lower 3pt\hbox{$\mathchar"218$}}       
 \raise 2.0pt\hbox{$\mathchar"13C$}}}       
\def\gtapprox{\mathrel{\spose{\lower 3pt\hbox{$\mathchar"218$}}       
 \raise 2.0pt\hbox{$\mathchar"13E$}}}       
\def\inapprox{\mathrel{\spose{\lower 3pt\hbox{$\mathchar"218$}}       
 \raise 2.0pt\hbox{$\mathchar"232$}}}

\begin{document}  
\title{Reply to the comment on ``On the problem of initial conditions in  
cosmological N-body simulations''}   
\author{Thierry Baertschiger \inst{1} and Francesco Sylos 
Labini\inst{2}}   
\institute{   
  \inst{1} D\'epartement ~de Physique Th\'eorique,   
  Universit\'e de Gen\`eve 24, Quai E. Ansermet,    
  CH-1211 Gen\`eve, Switzerland \\   
  \inst{2}Laboratoire de Physique Th\'eorique,  
  Universit\'e Paris-XI, B\^atiment 211, F-91405 
  Orsay Cedex, France \\  }    
\maketitle   
 
In \cite{alvaro} the authors criticize our measurements 
\cite{bsl02} of the reduced 
two-point correlation function $\xi(r)$ and of the one-point number 
variance $\sigma^2(r)$.  On one hand, they 
confirm our conclusion  that the estimator of $\xi(r)$  
for the initial conditions (IC) is in {\it disagreement} with the  
theoretical prediction at any scale in the simulation. 
However, the authors ascribe such a disagreement to the estimator's noise 
which they claim to   
be larger than the expected signal at any scale in the  
simulation. 
On the other hand,  they find that the behavior of the estimator of 
$\sigma^2(r)$ 
is in good agreement with the theoretical expected behavior. 
Note that they do not provide any argument to explain  
the mismatch between the behaviours of $\xi(r)$ and $\sigma^2(r)$. 
In order to clarify the matter we have firstly recomputed  
both estimators and we have increased the statistical  
precision. Our result is that the new estimation of $\xi(r)$  
is in agreement with our previous one and with the one by 
\cite{alvaro},  
while the new $\sigma^2(r)$ is in a better agreement with the  
theoretical behavior (see Fig.1).  
Hence, at least from the point of view of 
estimations, our conclusions are in good agreement  
with those of \cite{alvaro}. 
Our previous determination of  $\sigma^2(r)$ was affected 
by an imprecision in the normalization of the box 
which affected our estimation at scales  
comparable to the box size. 
The main point of disagreement with  
\cite{alvaro} 
concerns the interpretation of the  
measured statistical estimators, and their relation with the  
expected theoretical behaviors. The following discussion  
will allow us to clarify also the  mismatch in the  
determinations of  $\sigma^2(r)$.

The estimators  
of $\sigma^2(r)$ and $\xi(r)$ are directly related  
(fluctuations in number of points in balls or shells 
centered on a random or distribution point normalized to the mean 
density \cite{hz})  
and  
there is a precise theoretical relation with links these  
two statistics for a distribution with average density  
$n_0$:  
\be 
\label{eq1} 
\sigma^2(r) =  \frac{1}{V(r)^2} \int_V d^3x\int_Vd^3y\, 
\tilde \xi(|\vec{x}-\vec{y}|) =  
\frac{1}{n_0V(r)}+ \frac{1}{V(r)^2} \int_V d^3x\int_Vd^3y\, 
\xi(|\vec{x}-\vec{y}|) \,, 
\ee 
where  
the second equality  
is explicitly satisfied by {\it any} discrete distribution;   
we have defined  $\xi(r)$ to be the non diagonal part 
of the complete correlation function  $\tilde \xi(r) = \delta(r)/n_0 + 
\xi(r)$ and  hereafter we refer to the first term  
on the right hand side of Eq.\ref{eq1} as the ``discrete term'' 
while the second part is denominated ``correlated term''. 
Note that in the simulation we estimate  
the non diagonal part of the correlation function  
$\xi_S(r)$ (while the theoretical model gives 
the prediction for the complete  $\tilde \xi_T(r)$ of a continuous field),  
and the estimator of $\sigma^2(r)$ contains both  
the discrete and correlated term (see \cite{hz} for more details).  
Now  
it is clear that, if  
the variance in the simulation $\sigma_S^2(r)$  
agrees with the  
theoretical one $\sigma_T^2(r)$ 
we expect that $\tilde \xi_T(r) \simeq \xi_S(r)$ {\it if}  
the discrete term in Eq.\ref{eq1}. 
is negligible with respect to the correlated term. 
However this is not the case: 
the contribution of $1/n_0V$ is of order $\sigma^2_T(r)$ 
at scales a few times larger than the mean interparticle distance  
$\langle \Lambda \rangle$ (see Fig.1). 
\begin{figure}
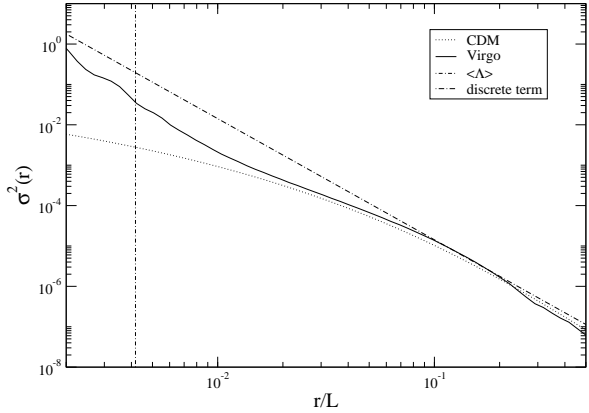
 
\onefigure[scale=0.32,angle=0]{sigmanew.eps} 
\caption{Normalized mass variance for the IC (Virgo),  
theoretical behavior (CDM) and  
the contribution of the ``discrete term''. 
The average distance between nearest neighbors  
$\langle \Lambda \rangle$  
is also reported. } 
\label{f} 
\end{figure} 
Hence our conclusion is that the  
agreement of the  
estimator of $\sigma^2(r)$ with $\sigma_T^2(r)$ is due to the fact that  
the particle distribution  
has still an average density $n_0$ too small  
to have a mass variance dominated by the  
``correlated term'' and discreteness effects are  
dominant at scale larger than 1/10th of the box size.

This is  
in agreement  
with the result that the 
behavior of the estimated  $\xi_S(r)$ is different  
at all scales from the expected one of $\tilde \xi_T(r)$: 
Its amplitude is very small on large scales  
and it becomes negligible with respect to  discrete noise. 
A simple estimation of such a  noise, by taking the {\it lower limit} 
of the Poisson case (i.e. the discrete term only), 
gives that  
$\delta \xi 
\approx 1/\sqrt{n_0} \approx  
1/\sqrt{256^3} \approx 2 \times 10^{-4}$ 
(as $V=1$ in our units).  
Using $\tilde \xi_T(r)$  
one can see that $|\tilde \xi_T| < \delta\xi$  
for $r\gtapprox 0.05  
\approx 10 \langle \Lambda \rangle$. Now, given the fact that at scales 
smaller than  
$ 3 \langle \Lambda \rangle \approx  0.015$ $\xi_S$  
is still oscillating between 
positive and negative values, due to  
the in-print of the pre-initial 
distribution, at best one can recover  
$\tilde \xi_T$  
in the range $\approx [3,10] \times  
\langle \Lambda \rangle $, while on larger scales 
the ``discrete''  noise predominates. 
Clearly such an agreement at intermediate scales  
is verified if there is not, in addition to ``discrete''  noise,  
a {\it systematic difference}   between $\tilde \xi_T$ and $\xi_S$: 
instead this seems to be the case as $\xi_S \ne \tilde \xi_T$ at all scales.  
For the power spectrum (which we have not discussed in our paper) 
the situation is rather similar to the case of the mass variance 
as $P(k)k^3 \approx \sigma^2(r=1/k)$ (but see \cite{hz}). 
We believe that a more complete  
theoretical understanding  
of the important problem of generating particle 
distribution with given correlations must be addressed  
by computing the effect of the application 
of a correlated displacement field on a generic particle 
distribution: 
We will discuss such a point in a forthcoming paper.

\end{document}